%% file: SnowmassBook-TF07-Colliders.tex
\documentclass{tcibook}

\usepackage{titletoc}

\usepackage{fancyhea}
\usepackage{work}
\usepackage{bm}       %    enables bold math symbols  e.g.  \bm{\gamma}
\usepackage{graphicx}
\usepackage{multirow}
\usepackage{lineno}
\usepackage{threeparttable}
\usepackage[normalem]{ulem}
\usepackage{color}
\usepackage[colorlinks = true,
            linkcolor = blue,
            urlcolor  = blue,
            citecolor = blue,
            anchorcolor = blue]{hyperref}

\input workshopsymbols.tex     
\begin{document}

%%  uncomment this line to use line numbers in drafts:
%\linenumbers

\pagenumbering{roman}

\parindent=0pt
\parskip=8pt
\setlength{\evensidemargin}{0pt}
\setlength{\oddsidemargin}{0pt}
\setlength{\marginparsep}{0.0in}
\setlength{\marginparwidth}{0.0in}
\marginparpush=0pt

% The content begins here

\pagenumbering{arabic}

%\renewcommand{\chapname}{chap:intro_}
%\renewcommand{\chapterdir}{.}
%\renewcommand{\arraystretch}{1.25}
%\addtolength{\arraycolsep}{-3pt}

\input Theory/TF07/Colliders.tex

\end{document}

%% file: workshopsymbols.tex
\def\beq{\begin{equation}}
\def\eeq#1{\label{#1}\end{equation}}
\def\eeqn{\end{equation}}

\newenvironment{Eqnarray}%
   {\arraycolsep 0.14em\begin{eqnarray}}{\end{eqnarray}}
\def\beqa{\begin{Eqnarray}}
\def\eeqa#1{\label{#1}\end{Eqnarray}}
\def\eeqan{\end{Eqnarray}}

\let\bar=\overbar

\def\lsim{\mathrel{\raise.3ex\hbox{$<$\kern-.75em\lower1ex\hbox{$\sim$}}}}
\def\gsim{\mathrel{\raise.3ex\hbox{$>$\kern-.75em\lower1ex\hbox{$\sim$}}}}

\def\del{\partial}
\def\Dslash{\not{\hbox{\kern-4pt $D$}}}
\def\dslash{\not{\hbox{\kern-2pt $\del$}}}
\def\pslash{\not{\hbox{\kern-2pt $p$}}}
\def\ETmiss{\not{\hbox{\kern-4pt $E$}}_T}

\def\Dlr{\mathrel{\raise1.5ex\hbox{$\leftrightarrow$\kern-1em\lower1.5ex\hbox{$D$}}}}

\def\MSB{{\bar{M \kern -2pt S}}}
\def\msb{{\bar{\scriptsize M \kern -1pt S}}}

\def\drb{{\bar{\scriptsize D \kern -1pt R}}}

%% file: Theory/TF07/Colliders.tex
\setcounter{chapter}{6} 

%% IMPORTANT:   from this file, refer to the bibliography as Theory/TF07/bibliography.tex   
%%    refer to a figure   A.pdf  as   Theory/TF07/figures/A.pdf  .
%\newcommand{\Shufang}[1]{\textcolor{red}{[SS: {#1}]}}
%\newcommand{\Fabio}[1]{\textcolor{green}{[FM: {#1}]}}
\newcommand{\jdt}[1]{\textcolor{blue}{[JDT: {#1}]}}

\DeclareRobustCommand{\Sec}[1]{Sec.~\ref{sec:#1}}
\DeclareRobustCommand{\Secs}[2]{Secs.~\ref{sec:#1} and \ref{sec:#2}}
\DeclareRobustCommand{\App}[1]{App.~\ref{app:#1}}
\DeclareRobustCommand{\Tab}[1]{Table~\ref{tab:#1}}
\DeclareRobustCommand{\Tabs}[2]{Tables~\ref{tab:#1} and \ref{tab:#2}}
\DeclareRobustCommand{\Fig}[1]{Fig.~\ref{ref:#1}}
\DeclareRobustCommand{\Figs}[2]{Figs.~\ref{ref:#1} and \ref{ref:#2}}
\DeclareRobustCommand{\Figss}[3]{Figs.~\ref{ref:#1}, \ref{ref:#2} and \ref{ref:#3}}
\DeclareRobustCommand{\Eq}[1]{Eq.~(\ref{eq:#1})}
\DeclareRobustCommand{\Eqs}[2]{Eqs.~(\ref{eq:#1}) and (\ref{eq:#2})}
\DeclareRobustCommand{\Ref}[1]{Ref.~\cite{#1}}
\DeclareRobustCommand{\Refs}[1]{Refs.~\cite{#1}}

\chapter{Theory of Collider Phenomena}

% abstract
\vspace*{0.3cm}
\begin{center}{{\Large \textsc{Abstract}}\\
\vspace*{0.3cm}
\begin{minipage}{5in}
Theoretical research has long played an essential role in interpreting data from high-energy particle colliders and motivating new accelerators to advance the energy and precision frontiers.
Collider phenomenology is an essential interface between theoretical models and experimental observations, since theoretical studies inspire experimental analyses while experimental results sharpen theoretical ideas.
This report---from the Snowmass 2021 Theory Frontier topical group for Collider Phenomenology (TF07)---showcases the dynamism, engagement, and motivations of collider phenomenologists by exposing selected exciting new directions and establishing key connections between cutting-edge theoretical advances and current and future experimental opportunities.
By investing in collider phenomenology, the high-energy physics community can help ensure that theoretical advances are translated into concrete tools that enable and enhance current and future experiments, and in turn, experimental results feed into a more complete theoretical understanding and motivate new questions and explorations.
\end{minipage}
}\end{center}

\vspace*{-0.0cm}
\begin{center}{{\Large \textsc{Table of Contents}}\\
\vspace*{0.3cm}
\startcontents[chapters]
\printcontents[chapters]{}{1}{}
}\end{center}

% conveners
\vspace*{-0.0cm}
\begin{center}{{\Large \textsc{Topical Group Conveners}}\\ 
\vspace*{0.2cm}
F. Maltoni$^{1,2}$,
S. Su$^{3}$,
J. Thaler$^{4,5,6}$\\
\vspace*{0.2cm}
\texttt{
\href{mailto:fabio.maltoni@uclouvain.be,shufang@arizona.edu,jthaler@mit.edu?subject=TF07 Snowmass Report}{fabio.maltoni@uclouvain.be, shufang@arizona.edu, jthaler@mit.edu}
}
}\end{center}
% authors 

\clearpage

\begin{center}{{\Large \textsc{Contributors}}\\
\vspace*{0.2cm}
T.~K.~Aarrestad$^{7}$,
A.~Aboubrahim$^{8}$,
S.~Adhikari$^{9,10}$,
I.~Agapov$^{11}$,
K.~Agashe$^{12}$,   
P.~Agrawal$^{13}$, 
S.~Airen$^{12}$, 
S.~Alioli$^{14}$, 
C.~A.~Arg\"{u}elles$^{15}$, 
Y.~Bao$^{16}$
% ...
}\end{center}

\begin{center}{\it\footnotesize
$^{1}$ Center for Cosmology, Particle Physics and Phenomenology, Universit\'{e} catholique de Louvain, B-1348 Louvain-la-Neuve, Belgium\\
$^{2}$Dipartimento di Fisica e Astronomia, Universit\`{a} di Bologna, Bologna, Italy\\
$^{3}$ Department of Physics, University of Arizona, Tucson, AZ 85721, USA\\
$^{4}$ Center for Theoretical Physics, Massachusetts Institute of Technology, Cambridge, MA 02139\\
$^{5}$ Co-Design Center for Quantum Advantage\\
$^{6}$ The NSF AI Institute for Artificial Intelligence and Fundamental Interactions\\
$^{7}$European Organization for Nuclear Research (CERN) CH-1211 Geneva 23, Switzerland\\
$^{8}$Institut f\"{u}r Theoretische Physik, Westf\"{a}lische Wilhelms-Universit\"{a}t M\"{u}nster,
$^{9}$Department of Physics and Astronomy, University of Kansas, Lawrence, Kansas 66045 U.S.A.\\
$^{10}$Department of Radiation Oncology and Molecular Radiation Sciences, School of Medicine, Johns Hopkins University, Baltimore, Maryland 21231, USA\\
$^{11}$Deutsches Elektronen-Synchrotron (DESY), Notkestraße 85, 22607 Hamburg, Germany\\
$^{12}$Maryland Center for Fundamental Physics, Department of Physics, University of Maryland, College Park, MD 20742, USA\\
$^{13}$Rudolf Peierls Centre for Theoretical Physics, University of Oxford, Parks Road, Oxford OX1 3PU, UK\\
$^{14}$Dipartimento di Fisica “G. Occhialini”, Universit\`{a} degli Studi di Milano-Bicocca, and
INFN, Sezione di Milano Bicocca, Piazza della Scienza 3, I – 20126 Milano, Italy\\
$^{15}$Department of Physics \& Laboratory for Particle Physics and Cosmology, Harvard University, Cambridge, MA\\
$^{16}$Department of Physics, Brown University, Providence, RI 02912, USA\\
% ...
}\end{center}

\newpage

\section{Introduction}

Theoretical research has long played an essential role in interpreting data from high-energy particle colliders and motivating new accelerators to advance the energy and precision frontiers.
The half-century quest to understand the Brout-Englert-Higgs mechanism---which describes the origin of elementary particle masses---is an inspiring, intertwined story of theoretical and experimental ingenuity.
Researchers in \textit{collider phenomenology} helped translate the quantum field theory that describes spontaneous electroweak symmetry breaking into the language of theoretical cross section predictions and experimental measurement strategies, culminating in the discovery of the Higgs boson at the LHC in 2012.
Today, precision studies of the Higgs boson, as well as broader investigations into physics beyond the Standard Model, are again being driven by advances in collider phenomenology.

In the coming decades, we foresee many opportunities to introduce novel \textit{observables}, \textit{calculations}, \textit{event generators}, \textit{interpretation tools}, and \textit{search strategies} to maximize the physics potential of the High-Luminosity LHC, advocate for new collider programs, and capitalize on future theoretical and experimental advances.
Collider phenomenology is an essential interface between theoretical models and experimental observations, since theoretical studies inspire experimental analyses while experimental results sharpen theoretical ideas.
By investing in collider phenomenology, the high-energy physics community can help ensure that theoretical advances are translated into concrete tools that enable and enhance current and future experiments, and in turn, experimental results feed into a more complete theoretical understanding and motivate new questions and explorations.

In this chapter, we showcase the dynamism, engagement, and motivations of collider phenomenologists by exposing selected exciting new directions and establishing key connections between cutting-edge theoretical advances and current and future experimental opportunities.
We also identify the most promising avenues where we expect major theoretical breakthroughs to take place in the coming years that could lead to transformative concepts and techniques in collider phenomenology. 

The rest of this chapter is organized as follows.
We begin in \Sec{tf07_summary} with a brief review of the role of collider phenomenology in high-energy physics research.
We then describe recent and anticipated advances in collider phenomenology, organized around key themes: 
observables (\Sec{tf07_observables}), 
calculations (\Sec{tf07_calculations}), 
event generators (\Sec{tf07_generators}), 
interpretation tools (\Sec{tf07_interpretation}), 
and search strategies (\Sec{tf07_search}).
We highlight theoretical studies of future colliders in \Sec{tf07_future}, and we conclude in \Sec{tf07_executive} with an executive summary.

\section{The Role of Collider Phenomenology}
\label{sec:tf07_summary}

At a fundamental level, all theoretical predictions for collider physics boil down to computing cross sections.
In schematic form, a cross section can be written as
\begin{equation}
\label{eq:cross_section}
\sigma_{\rm obs} \simeq \frac{1}{2 E_{\rm CM}^2}\sum_{n=2}^{\infty} \int \mathrm{d}\Phi_n \, |\mathcal{M}_{AB \to 12 \ldots n}|^2 f_{\rm obs}(\Phi_n).
\end{equation}
This formula contains a number of ingredients:
\begin{itemize}
    \item \textbf{Beam information}:  The center of mass collision energy $E_{\rm CM}$ and beam types $A$ and $B$.  For the high luminosity LHC (HL-LHC), $E_{\rm CM} = 14~\mathrm{TeV}$ and the beams are both protons.
    A variety of different beam types are being considered for future colliders (\Sec{tf07_future}).
    \item \textbf{Sum over final-state multiplicity}:  Since there is always a quantum probability to emit an arbitrarily soft particle, one has to account for the complete sum over $n$ for cross sections to be well-defined.  In the case that final-state particles are stable, the sum starts at $n = 2$.
    \item \textbf{Integration over phase space}:  Even though particle detectors have finite acceptance, theoretical calculations are based on integration over the full Lorentz-invariant $n$-body phase space $\Phi_n$.
    Monte Carlo event generators (\Sec{tf07_generators}) are a ubiquitous tool to perform this integration, as this approach automatically allows to obtain a fully flexible physics simulation of the scattering process.
    \item \textbf{Scattering amplitude}:  The quantum scattering amplitude $\mathcal{M}_{AB \to 12 \ldots n}$ is a complex object that cannot usually be computed from first principles, though in many cases one can factorize the squared-amplitude $|\mathcal{M}_{AB \to 12 \ldots n}|^2$ into a product of perturbative and non-perturbative terms.  In principle, this amplitude could be derived from the SM Lagrangian or come from novel scenarios beyond the SM (\Sec{tf07_search}).
    In practice, non-perturbative inputs are obtained directly from data or by ab initio quantum field calculations, such as those performed on the lattice, while the perturbative parts are organised in terms of a power expansion in the couplings. In  either case, the ability to perform calculations (\Sec{tf07_calculations}) is the key to making predictions whose final accuracy is set by the experimental needs. 
    \item \textbf{Observable}:  The function $f_{\rm obs}$ determines which regions of phase space contribute to a cross section.  The goal is to pick observables that can be predicted accurately in theory and measured as accurately in experiment, while probing the desired physics of interest (\Sec{tf07_observables}).
\end{itemize}
Finally, once one has computed a (fully differential) cross section, there is the key question of:
\begin{itemize}
  \item \textbf{Interpretation}:  How does one compare theoretical calculations of $\sigma_{\rm obs}$ to experimental cross section measurements and thereby draw conclusions about the structure of the universe?
  There are various interpretation tools (\Sec{tf07_interpretation}) depending on one's physics goals.
\end{itemize}

The field of collider phenomenology is aimed at linking theoretical models to experimental measurements and vice versa, by innovating each element of \Eq{cross_section} to form a bridge between the two communities.
Two recent developments have blurred the lines between theory and experiment, leading to exciting conversations across subfields:
\begin{itemize}
    \item \textbf{The rise of deep learning}:  Artificial intelligence and machine learning (AI/ML) have shown great promise to maximize the discovery potential of collider experiments.
    While one can use AI/ML to improve theoretical predictions and experimental protocols separately, some of the most powerful advances involve data-driven approaches where theoretical and experimental insights both play a role.
    \item \textbf{The availability of public collider data}:  Starting in 2014, the CMS experiment at the LHC released research-grade collider data for unrestricted use.
    This opened the door for proof-of-concept studies by theoretical physicists outside of the experimental collaborations.
    It has also highlighted the need for archival data strategies since theoretical progress can continue well after an experiment is decommissioned.
\end{itemize}
At the same time, experimental advances have highlighted the need for purely theoretical developments to advance the field:
\begin{itemize}
    \item \textbf{The need for precision}:  An increasing number of flagship experimental measurements are becoming limited by theoretical uncertainties, so properly interpreting experimental data in the context of underlying physical models requires improvements to our theoretical understanding.
    \item \textbf{Concrete targets for future colliders}:  There are a number of proposals for future colliders after the HL-LHC, including precision Higgs factories and high-energy discovery machines (\Sec{tf07_future}).
    Collider phenomenology is an essential tool to estimate the physics reach of these machines and define benchmarks to shape the machine and detector designs, which feeds into broader conversations about how best to invest HEP resources.
\end{itemize}

In the following sections, we highlight some exciting developments in each of these areas.
Note that there is substantial overlap between this report and other developments reported elsewhere in the Theory and Energy Frontiers. We invite the reader to consult the corresponding reports.

\section{Observables}
\label{sec:tf07_observables}

%%%%%
% Preamble paragraph
%%%%%

At the heart of collider phenomenology is choosing and defining observables.
The most basic observable is the total cross section.
It represents the simplest, most robust observable that can be measured and predicted.
Yet, it also contains the least information about   the microscopic physics of interest.
In general, the challenge of designing good observables stems from the tension between increasing the information content while keeping predictability and robustness against theoretical as well as experimental uncertainties. 
For example, by making more differential measurements, one gains access to different regions of phase space and rarer processes, and therefore increases the information. 
The design of observables also plays a key role in separating out known background processes from the signals of interest.

%%%%%
% Kinematic features paragraph
%%%%%

For many collider applications, it is important to develop observables that expose \textbf{kinematic features}.
Lorentz-invariant $n$-body phase space has rich structures, especially if there are intermediate resonances in long decay chains.
Specialized observables have been developed to isolate these kinematic features \cite{Franceschini:2022vck}, and we expect them to play an important role in studying new physics scenarios with multi-body final states~\cite{Chekanov:2021huv}.
For unpolarized decays, there is a fascinating ``invariance'' associated with the peak of the energy of the decay products~\cite{Agashe:2022sxw}, which has been exploited to study top quark decays.

%%%%%
% Jet substructure paragraph
%%%%%

Over the past decade, there has been enormous progress in the field of \textbf{jet substructure}~\cite{Nachman:2022emq}.
Jets are collimated sprays of particles that arise from the showering and fragmentation of high-energy quarks and gluons.
Collimated jets can also arise if heavier objects, such as top quarks or electroweak bosons, are produced with a large Lorentz boost.
In the past, a key application of jet substructure has been for \emph{tagging} jets arising from Standard Model partons.
Going forward, we anticipate progress in tagging exotic jet objects arising from new physics scenarios.
For example, long-lived Lorentz-boosted resonances can give rise to \emph{displaced fat jet} signatures~\cite{Padhan:2022fak}.
Confining dark sectors can give rise to \emph{semi-visible jets}, which require new tagging strategies~\cite{Beauchesne:2021qrw}.

%%%%%
% Correlator paragraph
%%%%%

For precision studies of collider data, there has been a revival of interest in \textbf{multi-point energy correlators}.
Multi-point correlators are still described by the theoretical formalism in \Eq{cross_section}, but they require a shift in thinking from the experimental perspective. 
In the case of standard observables, each collider event contributes one entry to a histogram.
For energy correlators, by contrast, each event contributes to multiple histogram bins, with entries weighted by (products of) particle energies.
Because of this energy weighting, energy correlators are less sensitive to soft physics and can be computed to high accuracy~\cite{Neill:2022lqx}. 
In addition, one can leverage insights from the conformal limit to understand the scaling behavior of multi-point correlators.

%%%%%
% Machine learning paragraph
%%%%%

In addition to observables defined from first-principles consideration, there is an explosion of work on \textbf{machine-learning-based observables}, where data itself is used to accomplish specific analysis tasks.
For supervised learning, this requires reliable training data, and therefore investment in precision event generation tools (\Sec{tf07_generators}).
A key insight is that machine learning architectures can incorporate \emph{symmetry group equivariance}~\cite{Bogatskiy:2022hub}, such that these methods respect the know theoretical structures of collider physics, such as permutation and Lorentz symmetries.
The continued rise of machine learning will require increased attention to the computational costs of these methods~\cite{Harris:2022qtm}.

%%%%%
% Optimal transport paragraph
%%%%%

A key example of an unsupervised learning strategy for collider phenomenology is \textbf{optimal transport}.
Unlike in \Eq{cross_section}, where each collider event is treated independently, optimal transport is based on relationships between pairs of collider events.
This open new geometric strategies for collider data analysis \cite{Komiske:2019fks}, including quantifying the dimensionality of the space of events.
Ideas from optimal transport have been leveraged to develop observables like \emph{event isotropy}, which are relevant for probing strongly-coupled new physics scenarios~\cite{Agrawal:2022rqd}.

%%%%%
% Quantum computing paragraph
%%%%%

With the rise of more sophisticated data analysis techniques, there has been recent interest in whether \textbf{quantum algorithms} might be relevant for high-energy data analysis~\cite{Delgado:2022tpc}.
Quantum algorithms have been successfully developed to compute specific collider observables, particularly those related to jet physics.
Whether these algorithms might yield a quantum advantage depends on what quantum hardware will eventually be available.
In particular, many quantum algorithms face significant computational overhead from encoding classical data into quantum form.
Dramatic computational gains could be possible if this encoding could be made more efficient.

\section{Calculations}
\label{sec:tf07_calculations}

Once an observable has been defined, the theoretical priority is predicting its distribution accurately enough in the Standard Model to match current and expected precision from experimental measurements.
In addition, predictions in specific beyond the Standard Model (BSM) scenarios or in effective field theories are needed to establish the sensitivity to new physics as well as to devise optimal search strategies.

Recent process in {perturbative calculations}---where predictions for observables are obtained order by order in small coupling expansions---has been tremendous.
This has been mostly due to a cross-fertilization between formal/mathematical developments in understanding the properties of multi-loop amplitudes and to a clever and efficient reorganization of different contributions into infrared (IR) safe observables up to next-to-next-to-leading order (NNLO) \cite{Cordero:2022gsh}.
Fixed-order perturbative computations of cross sections, however, often suffer from large logarithmic corrections, which must be resummed to all orders to restore the reliability of predictions from first principles.
Significant progress in devising new strategies and formalisms to obtained full {resummed results} has also been achieved.
Finally, applications of combining fixed order calculations with resummed ones, including those in Monte Carlo generators, to complement their accuracy in different areas of the phase space, have become the standard. 
Large room for improvement remains for the non-perturbative inputs.
Recent proposals for new approaches in quantum field theory---from measurement of pseudo-parton distribution functions to the extraction of the strong coupling from lattice---offer novel promising avenues to go beyond phenomenological models.

%%%%%
% Higgs/top physics
%%%%%

One of the key goals of future collider experiments (see \Sec{tf07_future}) is the precise study of the electroweak (EW) scale and in particular \textbf{precision EW, Higgs, and top physics}.
For example, Drell-Yan as well as Higgs boson production in $pp$ collisions are now  at the  N$^3$LO (QCD) + NLO (EW) + NLO (mixed EW/QCD) level.
Top/antitop production is at NNLO (QCD) + NLO (EW), moving now the first exploratory steps in the N$^3$LO region.
Further developments to reach N$^3$LO accuracy also at the differential level are expected~\cite{Caola:2022ayt}.
These calculations are enabling in an essential way for precision Higgs physics, which is central experimental program for the LHC and a possible future Higgs factory.

Beyond fixed-order results, significant improvements are expected by novel developments in \textbf{resummation techniques}, which are essential for high-precision phenomenology at current and future high-energy collider experiments.
Particular developments include resummation beyond leading power, the joint resummation of different classes of logarithms relevant for jets and their substructure, small-$x$ resummation in the high-energy regime, and the QCD fragmentation processes \cite{vanBeekveld:2022blq}.
Key to these developments is progress on understanding the factorization structure of gauge theories, including collinear factorization in QCD~\cite{Sterman:2022gyf}.
Factorization allows for scattering cross sections to be systematically decomposed into different (and sometimes universal) components that describe physics at different scales in the problem.

The corresponding situation at $e^+e^-$ collider is quite different, with large room for improvements in \textbf{precision calculations for lepton colliders}.
To date, no fixed-order results at two-loop accuracy are available for non-resonant processes, not even for the most important process at Higgs factories, i.e., $e^{+} e^{-} \to HZ$.
Similarly, {resummation of QED radiation} at the accuracy needed to match the expected accuracy of the experiments offers many compelling challenges \cite{Frixione:2022ofv, Jadach:2019bye,Freitas:2019bre}.

At the phenomenological level, new opportunities will open up at future colliders and \textbf{studies of elusive signatures} must be continuously explored.
For example, production of multi-boson final states \cite{Han:2021lnp} or very rare inclusive and semi-inclusive Higgs boson decay modes \cite{Han:2022rwq} might bring up new opportunities to access light-quark Higgs couplings.

Proven as well as novel methods have been put forward for \textbf{extracting Standard Model parameters}.
For example, the top-quark mass is a fundamental parameter of the Standard Model, relevant for understanding the stability of the electroweak scale.
The top mass could be extracted from the comparison of theory predictions and experimental measurements of differential cross-sections in $t\bar t$ final states at hadron colliders \cite{Alioli:2022ttk} or at the $t \bar t$ threshold in $\mu^+\mu^-$ collisions \cite{Franceschini:2022veh}.

As collider energies rise above the mass scales of the $W$, $Z$, and Higgs bosons, \textbf{electroweak radiation} becomes an increasingly important effect.
The very high energies achievable at a 100 TeV $pp$ collider and a multi-TeV $\mu^+\mu^-$ collider will enable probes of a new regime of EW interactions.
At these high scales, EW symmetry is effectively restored and the weak scale becomes small relative to the much larger collision energy.
In this regime, novel phenomena involving both initial and final state EW radiation will be triggered, with multiple collinear and/or soft emissions enhanced by large logs.
At such scales,  EW radiation displays similarities with QCD and QED radiation, yet with remarkable differences that make a full understanding of this regime an active line of theoretical research~\cite{Buarque:2021dji,Ruiz:2021tdt,Han:2022laq}.

\section{Event Generators}
\label{sec:tf07_generators}

Very few theoretical calculations can be performed analytically, so Monte Carlo event generators are the workhorse strategy to make  theoretical predictions which can most directly be compared to experimental data. 
Once these generators are interfaced with non-perturbative fragmentation models and parton distribution functions, the generator outputs can be used an inputs to experimental detector simulations.
The progress in event generators has mirrored the progress in calculations, with exciting developments in perturbative and resummed tools.

%%%%%
% Event generators in general
%%%%%

As experimental methods become more sophisticated, theoretical innovation in \textbf{multi-purpose event generators} becomes increasingly important~\cite{Campbell:2022qmc}.
In precision collider measurements, a significant (or even dominating) source of uncertainties of experimental analyses is often associated with event generators.
Those uncertainties can arise from different sources, such as the underlying physics models and theory, the truncation of perturbative expansions, the non-perturbative input and order of the evolution of the PDFs, the modeling and tuning of non-perturbative effects, and, finally, the uncertainty on the fundamental parameters of the theory.

A significant part of recent development activities in Monte Carlo event generators focuses on \textbf{extending their applicability and reducing their uncertainties}.
This depends critically on the collider types and experimental setups, especially for lower energy processes where non-perturbative effects dominate.
At next generation neutrino experiments---such as DUNE, HyperK, and SBN---the main systematic uncertainties to the measurements will arise from predicting and modeling neutrino-nucleus interactions.
At the Electron-Ion Collider, precise measurements of deep inelastic scattering (DIS) and other processes over the complete relevant kinematic range will be performed, advancing our understanding of hadronization as well as QCD factorization and evolution.
This will require the development, validation, and support of novel physics models.
The Forward Physics Facility at the LHC will leverage the intense beam of neutrinos, and possibly undiscovered particles, in the far-forward direction to search for new physics and calibrate forward particle production.
These measurements will require an improved description of forward heavy-flavor production, neutrino scattering in the TeV range, and hadronization inside nuclear matter.

Going to higher energies, future lepton colliders would provide per mille level measurements of fundamental parameters of the Standard Model, such as Higgs boson couplings and $W$ boson and top-quark masses.
The expected experimental precision will require event generators at an unprecedented accuracy at the fixed-order (EW and QCD) as well as in the resummation (QED and QCD).
At a multi-TeV lepton collider, such as a muon collider, a new range of phenomena involving EW initial-state and final-state radiation and possibly their resummation, which will be probed for the first time in a controlled environment.

From the theory point of view, event generation at present and future colliders feature many \textbf{common ingredients}:  higher-order QCD and EW corrections, factorization theorems and parton evolution, resummation of QCD and QED effects, hadronization, and initial and final-state modeling.
Several experiments will also need an improved understanding of heavy-ion collisions and nuclear dynamics at high energies.
These common challenges will benefit from {cross-talk in the Monte Carlo community}.
In addition to the physics components, there are also computing elements, such as interfaces to external tools for analysis, handling of tuning and systematics, and the need for improved computational efficiency.

%%%%%
% Computational cost
%%%%%

In addition to being a theoretical challenge, the increase in precision is becoming a \textbf{computational challenge}. \cite{HSFPhysicsEventGeneratorWG:2020gxw,HSFPhysicsEventGeneratorWG:2021xti}.
The reason is twofold.
First, the complexity of higher-order predictions for multi-particle final states increases exponentially with the order and number of final states.
Thus, the desired accuracy of the calculations, and the need of including also estimates for the uncertainties, make them increasingly expensive.
Second, the expected increase in collected data, for example at the HL-LHC, calls for the production of very large set of fully simulated events.
Among the goals of community, priority is given to: (i) understanding the best way to promote codes working on standard architecture to GPUs and vectorized codes; (ii) optimizing phase space sampling and integration algorithms, also by employing machine learning techniques such as neural networks, (iii) promoting research and development on how to eliminate or reduce the cost associated with negative weight events using new theoretical approaches; and (iv) supporting  collaboration, training, funding and career opportunities in the generator area.

%%%%%
% Machine learning
%%%%%

One way of address the ballooning computational cost of event generators is through \textbf{machine-learning-based generators}~\cite{Butter:2022rso}.
Modern machine learning is driving recent progress in event generation, simulation, and inference for high-energy colliders.
It is benefiting in a transformative way from new ideas, concepts, and tools developed under the broad umbrella of AI/ML research.
Concrete improvements in LHC event generation and simulation, as well as new ideas for LHC analysis and inference, are rapidly leading towards particle-physics-specific contributions to applied machine learning.
This in turn is inspiring a new generation of high-energy physics researchers bridging theory, experiment, and statistics.  

%{\color{blue} Summary here: Need for a paradigm change in the way that large scale simulations are performed in view of the size of the event sample and precision to be attained. Example: NN fast and light code for generating events on the fly instead of saving billions of file on disk. }

\section{Interpretation Tools}
\label{sec:tf07_interpretation}

Interpretation lies at the intersection of theoretical and experimental collider physics, and it is a fascinating conversation about which community should take primary responsibility for interpreting cross sections.
In cases where theoretical calculations are made available in the form of event generators, experimentalists can compare their results to generator outputs.
In cases where experimental results are made available in the form of unfolded measurements, theorists can compare their calculations to published data.
Ultimately, interpretation requires dialogue between communities, and collider phenomenology is a language to facilitate that dialogue.

%%%%%
% model dependent searches
%%%%%

Model-specific searches for new physics are a well-understood strategy for data interpretation (see \Sec{tf07_search}), but there is increasing interest in \textbf{anomaly detection} for collider physics.
The LHC Olympics 2020 challenge~\cite{Kasieczka:2021xcg} highlighted the promise of machine learning to discover hidden features in complex data sets.
On the other hand, this study also highlighted the need for the high-energy physics community to develop a rigorous theoretical definition of an ``anomaly''.

%%%%%
% eft interpretation paragraph
%%%%%

A quasi-model-independent approach to interpretation is the use of \textbf{effective field theories}.
If the impact of heavy new physics states can be captured by contact interactions involving Standard Model fields, then one can do a systematic expansion order by order in power counting.
In the Standard Model effective field theory (SMEFT), the impact of dimension-6 operators has been extensively studied, especially at future collider~\cite{deBlas:2022ofj}.
While naively the impact of dimension-8 operator are subdominant, new structures can appear at this order with unique phenomenological implications~\cite{Alioli:2022fng}, for example in precision angular distributions.

%%%%%
% analysis preservation paragraph
%%%%%

A key challenge facing collider physics is \textbf{data and analysis preservation}~\cite{Bailey:2022tdz}.
Data from colliders is unique, and sometimes new theoretical ideas are proposed well after an accelerator has been decommissioned.
There are multiple different levels of archival data, from raw detector outputs to legacy measurements, each of which requires different preservation strategies~\cite{Nachman:2022ltz}.
To maximize the scientific potential of archival collider data sets, theoretical physicists should actively participate in proposing and stress-testing archival data strategies~\cite{Bellis:2022onc}.

\section{Search Strategies}
\label{sec:tf07_search}

There is a huge range of possible scenarios beyond the Standard Model, and each scenario requires different observables to maximize signal acceptance and minimize background contamination.
Model specific searches remain the gold standard for the field, since they provide a well-defined statistical framework for setting limits (or announcing a discovery).
Model agnostic searches, though, are gaining traction, as it becomes increasingly possible to automate certain aspects of the search process.

New physics scenarios typically predict new particles in addition to the SM particle content.
These new particles have evaded the observations so far either because they are too heavy or too weakly coupled to be copiously produced, or because their signatures suffer from large SM backgrounds.
Other than indirect searches via the virtual effects of new physics in SM processes, the direct search for new particles provides unambiguous evidence for BSM.
While it is impossible to cover the numerous studies in the literature for new physics searches, we list a few highlights below.

TeV-scale scalars, fermions, or vector bosons typically have \textbf{cascade decay signatures}.
They could appear in models with an extended Higgs sector, composite Higgs models with vector-like fermions~\cite{Kling:2022jcd,Dermisek:2022kyh,Banerjee:2022xmu,Adhikari:2022yaa,Aboubrahim:2021ily,Robens:2022lkn}, or warped extra dimension models with Kaluza-Klein (KK) modes~\cite{Agrawal:2022rqd,Agashe:2022gbb}.
For heavy particles produced at high-energy colliders, they could leads to boosted objects with interesting substructure.
Modern event shape variables like event isotropy or a variety of jet substructure observables (see Sec.~\ref{sec:tf07_observables}) can be used to identify the signal over the SM backgrounds.

The study of {\bf dark sectors} has been a subject of interest in recent years, including dark showers~\cite{Albouy:2022cin}, portal matter~\cite{Rizzo:2022qan}, and multi-field scenarios~\cite{Dienes:2022zbh}.
Weakly-coupled dark sectors feature missing energy, displaced vertex, and long-lived particle signatures at colliders.
For strongly-coupled dark sectors, there are additional signatures such as emerging or semi-visible jets for QCD-like scenarios, and soft unclustered energy patterns or glueballs for non-QCD-like scenarios.
Event-level variables, deep neural networks, auto-encoder-based anomaly detection or better triggering algorithms can be used to identify these unusual signatures.

Finally, \textbf{low-mass scalars} are a key target for current and future colliders, including axion-like particles~\cite{DiLuzio:2020wdo, Bao:2022onq, Han:2022mzp,Feng:2022inv} and light scalars from extended Higgs sectors~\cite{Robens:2022erq}.
Collider searches complement searches at astronomical, dark matter, and nuclear physics experiments~\cite{DiLuzio:2020wdo}.
 
\section{Physics at Future Colliders}
\label{sec:tf07_future}

There have been tremendous efforts by the experimental and theory communities in planning and designing future generation colliders.
The goal of these studies are to sharpen our understanding of existing physics and explore the potential of precision and energy frontier machines to facilitate the next discovery.
An important role of collider phenomenology is to define the physics cases for future colliders and to present physics benchmarks to shape the machine parameters and detector design.

In the precision frontier, a future {\bf high precision Higgs, electroweak, and top factory} has been considered ~\cite{Agapov:2022bhm,Gao:2022lew,ILCInternationalDevelopmentTeam:2022izu}.
The Future Circular Collider (FCC) and Circular Electron Positron Collider (CEPC)~\cite{FCC:2018byv, Bernardi:2022hny, CEPCStudyGroup:2018ghi} are two circular $e^+e^-$ colliders with similar design and physics reach.
These would run at four different center-of-mass energies: at the $Z$-pole, at the $WW$ threshold, at the $ZH$ threshold, and at the $t\bar{t}$ threshold.
The luminosities range from about $ 10^{36}\ {\rm cm}^{-2} {\rm s}^{-1}$ per interaction point at the $Z$ pole, to about $10^{34}\ {\rm cm}^{-2} {\rm s}^{-1}$ per interaction point at the $t\bar{t}$ threshold.
Millions of Higgs bosons will be produced and most of the Higgs couplings can be measured at percent level, with $hZZ$ coupling at sub-percent level.
Precision measurements of electroweak and strong forces can be performed, which are sensitive to new physics contributions at loop level.
Rare process in heavy-flavor and tau physics can be probed beyond the HL-LHC reach.
In addition, these machines provide the best potential in the direct searches of dark matter, sterile neutrinos, and axion-like particles with masses up to about 90 GeV.

The International Linear Collider (ILC) is a {\bf linear $e^+e^-$ collider} with a starting center-of-mass energy at 250 GeV for Higgs physics.
It could accommodate a range of center-of-mass energies from $Z$ pole to 350 GeV ($t\bar{t}$) and above, up to 1 TeV~\cite{ILCInternationalDevelopmentTeam:2022izu, Baer:2013cma}.
In addition to the Higgs, electroweak, and top physics similar to the CEPC and FCC studies, higher center-of-mass energies allows the enhanced Higgs production rate via vector boson fusion  (VBF) process, and precise measurements of top Yukawa coupling and Higgs self couplings.
It also allows the direct detection of BSM particles, especially those with only electroweak interactions.
Beam polarization and complete reconstruction of final states also make the precise measurement of tri-gauge boson couplings possible.
Other physics potential includes the beam dump and dedicated fixed-target experiments to search for light weakly-interacting particles.
The Compact Linear Collier (CLIC) is a high energy, high luminosity $e^+e^-$ collider with center of mass energy at 380 GeV, 1.5 TeV, and 3 TeV~\cite{Brunner:2022usy, Roloff:2652257}, which offers a unique combination of high collision energy and clean environment.
It can probe new physics beyond the SM with both indirect search and direct production of heavy new particles.
The recently proposed Cool Copper Collider (${\rm C}^3$)~\cite{Bai:2021rdg, Dasu:2022nux} offers another possibility for the realization of a multi-TeV lepton machine in the future.

At the Energy frontier for {\bf hadron colliders}, the HL-LHC will extend the existing LHC program with $pp$ collision at $\sqrt{s}=14$ TeV and an integrated luminosity of 3 ${\rm ab}^{-1}$ \cite{CMS-PAS-FTR-22-001, Dainese:2703572}.
With twenty times the current available dataset, HL-LHC can perform precise measurements of Higgs, top properties, study rare process in the flavor sector, perform precision QCD measurements and heavy ion related analyses, as well as direct search of new particles beyond the SM at the current energy frontier.
Future hadron colliders under consideration are the FCC-hh~\cite{Benedikt:2022kan} and Super Proton Proton Collider (SPPC)~\cite{Tang:2022fzs}, which are 100 TeV $pp$ colliders that improve the heavy new particle reach of the LHC by another order of magnitude~\cite{Bernardi:2022hny}.
In addition, Higgs self-coupling could be measured with about 3 $-$ 8\% precision, which could help to pin down the shape of the Higgs potential.
Relevant parameter space of weakly-interacting massive particle scenarios of thermal relic dark matter can be tested.
Furthermore, QCD matters at high density and high temperature can be studied.
Using the existing LHC tunnel, a 27 TeV $pp$ collider with an integrated luminosity at least three times of the HL-LHC has also been considered~\cite{Zimmermann:2651305}.
 
{\bf Lepton-hadron colliders} have also been proposed to move the field of deep inelastic scattering (DIS) to the energy and intensity frontier of particle physics.
The FCC-eh~\cite{Bernardi:2022hny} collides 60 GeV electron beam with 50 TeV proton beam at the FCC tunnel with luminosity of $10^{33}-10^{35}\ {\rm cm}^{-2} {\rm s}^{-1}$.
It could provide precise information on the quark and gluon structure of the proton in an unprecedented kinematic reach, and a per mille accurate measurement of the strong coupling constant.
It also enables unique measurements of eletroweak parameters and flavor structure of the SM.
Direct searches on leptoquark as well as contact interactions up to tens of TeV scale can also be explored.
A low energy version of the Large Hadron-electron Collider (LHeC)~\cite{LHeC:2020van} with the LHC proton beam has also been considered.
 
An {\bf Electron Ion Collider} (EIC)~\cite{AbdulKhalek:2022erw} planned at Brookhaven National Laboratory will collider polarized electron beam with polarized protons or light ions with variable center of mass energy 20 $-$ 100 GeV and luminosity similar to that of LHeC and FCC-eh.
The science cases include beyond the SM new physics, tomogography of hadrons and nuclei, jet physics, heavy flavor physics, and physics at small Bjoeken-$x$, which will present many new opportunities to address some of the crucial and fundamental open scientific questions in particle physics.  

A {\bf Muon collider}~\cite{MuonCollider:2022xlm, Aime:2022flm} with center of mass energy ranging from 3 TeV to 10 TeV or even higher has been discussed, which provides a unique combination of high energy approach and precision measurements with a relatively compact collider ring.
For direct production of new particles at a muon collider, discovery up to the production kinematic threshold is typically possible.
A higher energy muon collider can also be viewed as a vector boson collider with the electroweak radiation of a nearly on-shell massive vector boson.
VBF production of both the SM Higgs and new particles will become dominating, especially at higher center-of-mass energy.
Precise measurements of Higgs couplings comparable to an $e^+e^-$ Higgs factory are possible, as well as accurate measurements of the Higgs trilinear coupling to a few percent level.  High-energy measurements of SM processes at a muon collider also makes it possible to be sensitive to new physics scale of about 100 TeV via indirect measurements.
Dark sector via Higgs portal, or dark matter can also be studied at a muon collider.
There are also muon specific opportunities like resolving new physics that is responsible for the muon $g-2$ anomaly or lepton flavor universality anomaly.
In addition, electroweak radiation at such a high energy region can be studied, which is  important for developing a correct physical picture as well as achieving the needed accuracy of the theoretical predictions.

A {\bf gamma-gamma collider}~\cite{Serbo:2005vu, Barklow:2022vkl, Barzi:2022rax} has been considered as a natural part of the linear collider proposals with high-energy photon beam generated through Compton back-scattering of laser light focused onto the incoming electron bunch just before the interaction point.
The physics case for a $\gamma \gamma$ collider with center-of-mass energy ranging from low energy to multi-TeV has been developed, including investigating the Higgs boson and its properties via single resonance production, searching for new particles beyond the SM, studying electroweak gauge boson physics, as well as probing photon structure and QCD physics.

The {\bf Forward Physics Facility} (FPF)~\cite{Anchordoqui:2021ghd,Feng:2022inv} is a proposal to create a cavern located in the far-forward region from the ATLAS interaction point at the LHC during the High Luminosity era.
It could host a suite of experiments that are designed to search for light and weak-interacting particles (FASER2) including long-lived particles (LLP), axion-like particles and dark sector particles, neutrinos at TeV energies (FASER$\nu$2 and Advanced SND), neutrinos and dark matter (FLArE), and millicharged particles (FORMOSA).
It will allow tests of BSM physics and probe our understanding of strong interactions,  proton and nuclear structure, neutrino physics at high energy, as well as astroparticle physics via the study of high-energy cosmic rays.
Another LLP search experiment,  Massive Timing Hodoscope for Ultra-Stable neutraL pArticles (MATHUSLA)~\cite{MATHUSLA:2022sze} situated at CERN on earth  surface near the CMS detector, would be able to extend the sensitivity in long lifetime and LLP cross section by several orders of magnitude compared to the main LHC detectors alone.
It can probe a wide range of LLP from GeV to TeV scale, as well as dark matter scenarios with dark matter from LLP decay. 
 
There are also studies on the {\bf Fermilab Booster Replacement}~\cite{Arrington:2022pon}, which can be used for searches of dark sectors, neutrino physics, 
a slew of charged lepton flavor violation searches, and precision measurements, as well as
R\&D towards a muon accelerator at the energy frontier.
Super-Compact Dimuonium Spectroscopy Collider (DIMUS)~\cite{Fox:2022lld} proposed at Fermilab can produce copious $\mu^+\mu^-$ atoms at the production threshold, allowing for precision tests of QED and opening the door for searches for new physics coupled to the muon.   

Collider studies for particular models or collider signatures at current and future colliders are a continuing effort in the community.
White paper contributions can be found at \Refs{Agrawal:2022rqd,Robens:2022lkn,Han:2022mzp,Han:2022ubw,Costantini:2020stv,Barman:2022pip,Kalinowski:2022fot,Robens:2022erq,Han:2022edd,Kalinowski:2022fot,Papaefstathiou:2022oyi}, as well as those submitted to the Energy Frontier of Snowmass studies.

\section{Executive Summary}
\label{sec:tf07_executive}

Collider phenomenology is a highly active area of theoretical investigation that connects in crucial ways to the experimental energy and precision frontier program. 
Continued investment in this field is necessary to maximize the physics reach of the HL-LHC, to guide discussions about possible future collider programs, and then prepare for them. 
The phenomenology community has provided many inputs to the Snowmass activities, which are very briefly presented in this report. We would like to highlight the following points:

\begin{itemize}
\item Collider phenomenology is an \textbf{essential interface} between the theoretical and experimental high-energy physics communities. It brings together a broad range of expertise from different fields and serves various roles: from connecting formal investigations to experiments, such as providing guidance on the exploration of the physics possibilities, to supporting the experimental community with the essential tools to simulate and interpret data, bringing information back to the theory community.  
\item A \textbf{new generation of observables} are being developed that borrow from adjacent fields, such as optimal transport and conformal field theories. 
Because of their novel theoretical and experimental properties, they provide complementary probes of collider phenomena.
In some cases, these observables can be computed to higher accuracy than traditional observables.
In other cases, continued research is need to incorporate these techniques into the collider physics community.
\item \textbf{Precision calculations and state-of-the-art event generators} are essential to maximize the physics potential of the HL-LHC and future colliders.
Enormous progress has been achieved in the last two decades to improve predictions for $pp$ collisions. The corresponding conceptual and technical advances have not yet been applied to lepton colliders, leaving open the possibility for significant progress  in the coming years. 
As one goes to higher energies, effects like electroweak radiation must be included for sufficiently accurate theoretical descriptions.
\item To make accurate predictions in high-dimensional phase spaces, \textbf{new computational paradigms for theoretical calculations} must be pursued.
This is particularly important to enable the incorporation of theoretical uncertainties at reasonable computational costs.
\item There is a growing catalog of \textbf{model-agnostic search strategies} such as anomaly detection and effective field theory approaches, which complement well-established tools for model-specific searches.
\item \textbf{Exotic collider signatures}, such as long-lived particles and dark sector showers, are inspiring new data analysis strategies and motivate new experimental facilities.
\item In addition to searching for physics beyond the Standard Model, we have an opportunity to probe \textbf{unexplored Standard Model phenomena}.
This includes the non-perturbative dynamics of QCD and the restoration of electroweak symmetry at high energies.
\item \textbf{Machine learning} is having a transformational impact on collider phenomenology, from the way observables are defined, to the tools used to simulate collisions, to the strategies to hunt for anomalies. 
Continued advances will require interdisciplinary training in statistics, computation, and data science.
\item Collider data sets are unique, with legacy value long after an accelerator is decommissioned.
Theoretical physicists should continue to collaborate with the experimental community to advocate for and assist with \textbf{long-term data preservation efforts}.
One pathway towards long-term usability is making collider data public as a way to stress-test archival strategies and to encourage novel analyses by the theory community while the experimental collaborations are still active. 
\item Theoretical studies plays a key role in defining the \textbf{physics case for future colliders} and presenting physics benchmarks to shape machine parameters and detector design.
\item While the next collider after the LHC is at least one decade away, {\bf a widespread sense of urgency} permeates the collider phenomenology community.
Considerable work and a number of groundbreaking results are expected to fully exploit the potential of the data from the HL-LHC and future colliders. 
Developing the theoretical tools needed to meet experimental targets---and providing feedback to inform new and/or more precise theoretical frameworks---will require a community effort which can only be achieved with significant human and computing resources. 

\end{itemize}

\nocite{*}
\bibliographystyle{JHEP}
\bibliography{Theory/TF07/collider-references} 